# Sparsely repeated 21.7 Tb/s Net-Rate Transoceanic Transmission with 266 km Ultra-Long Spans Enabled by Low IMI and Low loss Hollow Core Fiber

**Rajiv Boddeda[1*], Carina Castineiras Carrero[1], Haïk Mardoyan[1], Amirhossein Ghazisaeidi[1], Peng Li[2], Shuhai Li[2], Lei Zhang[2], Jie Luo[2], and Jérémie Renaudier[1]**

[1]*Nokia Bell Labs, Optical Transmission Dept., Massy, France;*
[2]*State Key Laboratory of Optical Fiber and Cable Manufacture Technology, Yangtze Optical Fibre and Cable (YOFC), Wuhan, China.*
*\*rajiv.boddeda@nokia-bell-labs.com*

**Abstract:** We demonstrate 21.7-Tb/s net-rate transmission across 6660-km with 266-km ultra-long spans of HCF. By exploiting a newly designed GTA-ST-HCF, high-power booster, and adaptive channel rates, we realize WDM transoceanic transmission with fewer than 30 repeaters.

## 1. Introduction

The standard single-mode fibers (SMFs) have revolutionized how the information is transferred across the world and brought in an unprecedented change on how humans communicate [1]. Significant efforts are being made to boost the capacity of optical fiber communication systems ranging from digital signal processing (DSP) techniques to the design of optical fiber systems to continuously reduce the cost per bit [2-3]. More recently a new type of fiber known as the anti-resonant hollow core fiber (AR-HCF) is poised as a promising solution for next-generation optical transmission systems, by achieving ultra-low loss while overcoming the nonlinearity and latency limitations of SMFs at the same time [4]. Moreover, the lower dispersion of HCF helps limit the receiver's complexity as the baud-rate grows. Currently, long-distance optical transmission is limited by the inter-modal interference (IMI) and gas line absorption (GLA) in HCF where the interference with higher order modes impacts the signal quality [4,5,6].

In this paper, we report on a transoceanic WDM transmission with repeaters spaced by 266 km spans. Our WDM transmission setup is based on a recirculating loop made of one span of 266 km of a newly designed AR-HCF providing low loss and low IMI. Using a high-power Erbium-Ytterbium doped fiber amplifier operating in the C-band, we achieve successful transmission of 20+ Tb/s net data rate thanks to channel rate optimization to overcome the limitations imposed by GLA peaks at long distances. While standard transoceanic systems typically rely on 100+ repeaters, this result lays foundation for ultra-long-haul transmission systems with less than 30 repeaters, holding the promise for faster cable deployment and lower maintenance.

## 2. Experimental setup

Fig. 1 depicts the experimental set-up used for reproducing ultra-long-haul transmissions. The transmitted signal is made of a fully loaded wavelength division multiplexed (WDM) comb composed of carved amplified spontaneous emission (ASE) noise over 4.8 THz plus one channel under test (CUT). The CUT is made of a C-band tunable laser source (TLS) modulated with dual polarization 16-QAM signal at various symbol rates up to 135-GBaud. The digital waveforms are generated by digital-to-analog converters (DACs) from a commercially available arbitrary waveform generator operating at 256-GS/s with a 10-dB bandwidth of 70-GHz. The digital signal is Nyquist shaped (root-raised-cosine, roll-off 0.01) for all symbol rates. The generated WDM comb signal is sent into the recirculating loop and is first amplified using a carefully chosen 34 dBm amplifier with up to 35 dB gain and a maximum noise figure of 4.68 dB in the C-band. The amplified signal is launched at 34 dBm to the ultra-long span of 266 km of Gap Tube Assisted Support Tube-Hollow Core Fiber (GTA-ST-HCF). The 266 km span is terminated by SMF/HCF adaptors and is composed of 23 GTA-ST-HCF spools spliced together. As shown in the inset of Fig 1a, the GTA-ST-HCF combines a support tube architecture that enables ultra-low confinement loss with a gap tube assisted modal filtering mechanism, thereby achieving both low attenuation and high modal purity. Figure 1b presents the attenuation coefficient and dispersion as a function of wavelength over 266 km fiber at 1 pm resolution. One can observe the GLA absorption peaks at wavelengths between 1530-1545 nm. The GTA-ST-HCF structure effectively suppresses higher-order-mode coupling and reduces the IMI, which is highly advantageous for long-distance transmission [7]. Figure 1c illustrates the histogram of transmitted power for GTA-ST-HCF, measured by sweeping the laser wavelength from 1549 nm to 1551 nm using a narrow linewidth laser. For comparison, the typical probability density function (PDF) of HCF with an IMI of -52 dB/km [5] is overlaid in red. We observed a 6.4-fold reduction in the standard deviation of power fluctuations, leading to an estimated IMI of −68.8 dB/km. In Fig 2a, we show the bi-directional OTDR trace together with the loss profile estimated by subtracting the forward and backward traces; this profile is fitted with a linear function, yielding a slope of −0.098 dB/km at 1550 nm, corresponding to a span loss of about 27 dB at 1550 nm including the splicing losses and insertion losses from the SMF/HCF adaptors. At the end of the loop, the WDM comb signal is amplified and passed through a dynamic gain equalizer followed by polarization scrambler.

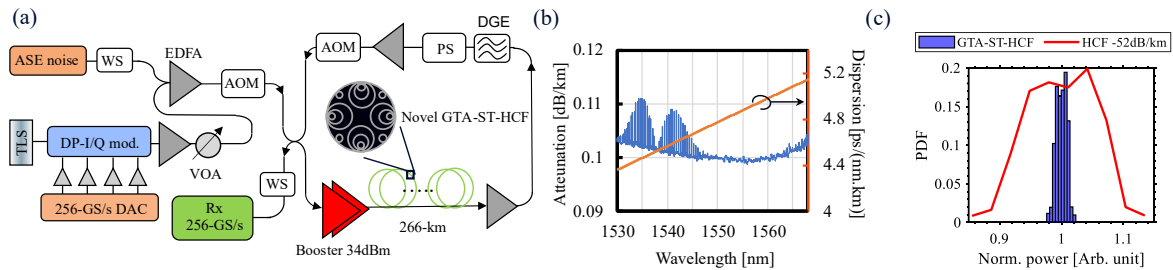

**Fig. 1**. a) Experimental setup and b) attenuation and dispersion of GTA-ST-HCF versus wavelength. TLS: Tunable Laser Source; WS: Wave Shaper; PS: polarization scrambler; AOM: acousto-optic modulator. VOA: Variable Optical Attenuator. EDFA: Erbium doped fiber amplifier. DGE: Dynamic Gain Equalizer. c) Histogram of measured transmitted power using sweep wavelength technique.

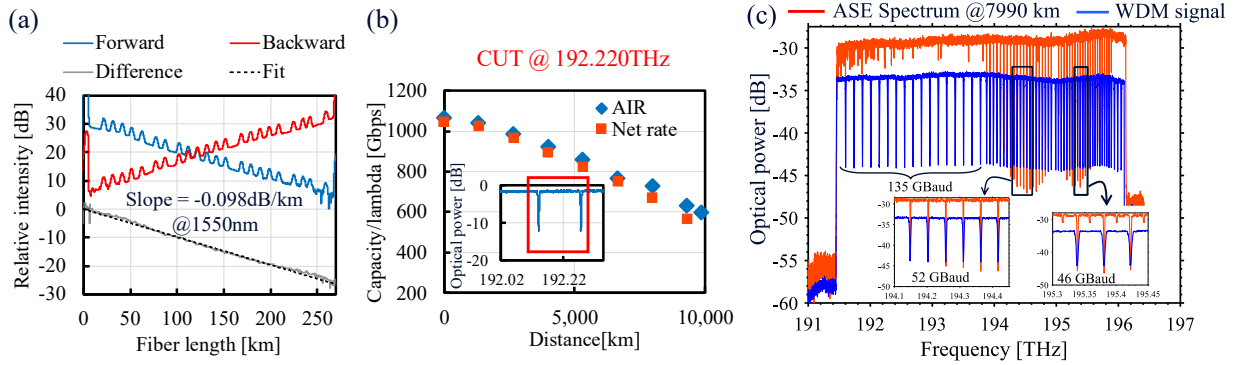

**Fig. 2**. a) Bi-directional OTDR trace of the 266-km span measured using 20 μs pulse. The difference between the forward and the backward trace is shown in grey along with a linear fit (dashed black) showing a loss coefficient of -0.098 dB/km. (b) The capacity per lambda as a function of distance. In the inset we show the signal spectrum around the CUT @ 192.220 THz (c) The ASE noise spectrum in red after 7990 km and the designed WDM channels to avoid the GLA, with insets showing 52 GBaud and 46 GBaud signals between GLA lines.

At the receiver side, after a given number of round trips inside the loop, the CUT is extracted from the WDM comb with a wave shaper configured as a bandpass filter. The CUT is then received by a standard coherent receiver front-end and sampled at 256-GS/s using a 110-GHz bandwidth real-time oscilloscope. For each measurement, a set of 5 waveforms is stored, each recorded waveform consisting of 8 million samples which are processed offline. The standard digital signal processing suite consists of matched filter, chromatic dispersion compensation, complex 2x2 MIMO adaptive equalization, frequency offset and phase compensation, and least-mean square equalizer to mitigate for transmitter I/Q imbalances. We finally compute the signal-to-noise ratio (SNR), the normalized generalized mutual information (NGMI) as well as the net-rate by applying multi-rate spatially coupled low-density parity check codes (SC-LDPC) with rates varying between 0.4 and 0.9 to achieve error-free decoding [8].

## 3. Transmission results

First, we assessed the performance of one channel located at 192.220 THz, outside the GLA peaks regions. We measured the performance of this channel modulated at 135 GBaud, packed in a 137.5 GHz grid. In Fig. 2b, we show the achievable information rate (AIR) and the net-rate measured as a function of the distance. We estimate an AIR reaching as high as 590 Gbps per lambda at 10000 km thanks to low repeater density and low NF capabilities of the 34 dBm amplifier. We limited our study to 16-QAM, since over distances up to 10,000 km the NGMI ranges from 0.9 down to 0.5, corresponding to code rates between 0.9 and 0.4. Using lower modulation formats like 4-QAM can extend the reach further but is beyond the scope of this study. Then, we captured the optical spectrum of ASE noise after 30 round trips i.e., 7990 km, where we observe the impact of GLA reaching as high as 18 dB as shown in red in Fig 2c.

To avoid GLA spectral notches while keeping spectral efficiency high, we adapt the channel baud rate (and therefore occupied bandwidth) per given spectral grid. We design the WDM signal by optimizing the symbol rate of the transmitted signal such that signal spectrum does not overlap GLA lines. In Fig 2c, we show the designed WDM comb signal. We employed symbol rates of 135 GBaud until almost 194 THz and then switched to 52 GBaud over the first band of GLA as shown in the left inset of Fig 2c. In the second band of GLA, we varied the symbol rates from 46 GBaud to 30 GBaud with spacing of 5-6 GHz as shown in the right inset of Fig 2c. In total, we have about 60 channels, of which 17 are at 135 GBaud. In Fig. 3a, we now show the measured SNR after 6660 km and 7990 km, corresponding to 25 and 30 round-trips. Thanks to the channel rate optimization with respect to the GLA regions, we achieved similar SNR between 191.5 THz and 195 THz. However, in the second band of GLA, we observed an additional penalty of 2-3 dB SNR which can be attributed to residual GLA lines as shown in the right inset of Fig. 2c. In Fig. 3c, we show the NGMI and the required code rate for error free decoding using SC-LDPC after 6660 km. The total AIR over the C-band is 25.8, 23.4, and 19.3 Tb/s, yielding net throughputs of 23.5, 21.7, and 15.9 Tb/s over 5320, 6660, and 7990 km, respectively. These values correspond to spectral efficiencies of 4.82, 4.48, and 3.3 b/s/Hz over the entire C-band despite the presence of GLA. Sustaining such high capacities over transoceanic distances, despite spectral impairments, highlights the effectiveness of adaptive baud rate signaling. In Fig. 3c, we show the state-of-the-art capacity times distance product achieved over one single core fiber using standard EDFAs. In single mode fibers, the record is at 570 Pbps.km, achieved using short spans of 55 km [8,9]. In hollow core fibers, using spans of 127 km, it has been shown very recently that reaching transoceanic distances is possible but challenging due to GLA/IMI induced impairments [5]. Here, we achieved a two-fold increase in the net throughput while using ultra-long spans which are more than two times longer. To further increase the capacity, entropy loading of multi-carriers enables a theoretically optimal water-filling optimization, albeit at the cost of increased transceiver complexity [11].

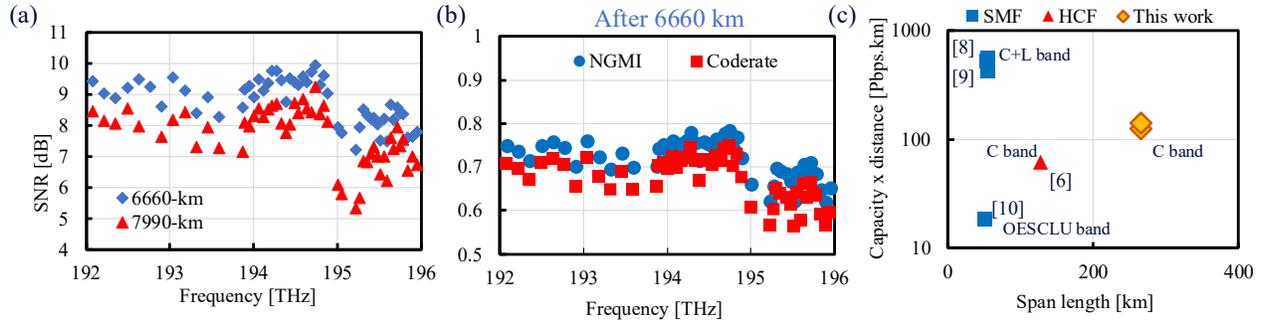

**Fig. 3**. a) The SNR of the received data as a function of frequency at 6660 km and 7990km. (b) The estimated NGMI and code rate required for error free decoding as a function of frequency after 6660 km. (c) The capacity times distance product as a function of span length.

## 4. Conclusion

To the best of our knowledge, this is the first demonstration of transoceanic WDM transmission using spans as long as 266 km. We achieved the transmission of total throughput of 21.7 Tb/s over 6660 km, corresponding to only 25 repeaters for a transatlantic distance. We also showed the net rate throughputs at other distances, reporting 23.5 Tbps over 5320 km and 15.9 Tbps over 7990 km. These pioneering results were demonstrated by leveraging newly designed GTA-ST-HCF with low loss and low IMI, high-power amplifier and variable baud rate spectral management to avoid GLA distortions while keeping spectral efficiency high. These results pave way for a paradigm shift in future optical networks in metro and submarine links where ultralong spans can bring up to a 5-fold reduction in repeaters.

**Acknowledgements:** The authors acknowledge Exfo/Infractive for lending the OTDR and Wave analyser equipment for the measurement.

## 5. References

[1] E. Agrell, M. Karlsson, F. Poletti, et al., "Roadmap on Optical Communications," J. Opt., vol. 26, 093001, pp. 1-64, 2024.
[2] B.J. Puttnam, et al., "Space-division Multiplexing for Optical Fiber Communications," Optica, vol. 8, no. 9, pp. 1186-1203, 2021.
[3] J. Renaudier, et al., "Devices and Fibers for Ultrawideband Optical Communications," IEEE, vol. 110, no. 11, pp. 1742 -1759, 2022.
[4] Y. Chen, et al., "Hollow Core DNANF Optical Fiber with <0.11 dB/km Loss," in OFC, San Diego, United States, paper Th4A.8, 2024.
[5] R. S. B. Ospina et al., "Leveraging Digital Subcarrier Multiplexing for Long-Haul Transmission over HCFs in the Presence of IMI", in OFC, Los Angeles, United States, paper Th1J.3, 2026.
[6] Y. Hong et al., "Real-Time Fully-Loaded C-band, Low-Latency, Long-Haul Transmission over Hollow-Core Fiber," 2025 European Conference on Optical Communications (ECOC), Copenhagen, Denmark, 2025, pp. 1-4.
[7] P. Li, et al., "Low Intermodal Interference and Low Loss Hollow Core Fibers", in OFC, Los Angeles, United States, paper M2J.1, 2026
[8] A. Ghazisaeidi, et al., "Advanced C+L Band Transoceanic Transmission Systems Based on Probabilistic Shaped PDM-64QAM", JLT, Vol. 35, no. 7, p. 1291 (2017).
[9] S. Zhang et al., "50.962 Tb/s over 11185 km bi-directional C+L transmission using optimized 32QAM," 2017 Conference on Lasers and Electro-Optics (CLEO), San Jose, CA, USA.
[10] B. J. Puttnam et al., "402 Tb/s GMI data-rate OESCLU-band transmission," presented at Proc. 2024 Opt. Fiber Commun. Conf. Exhibition San Diego, CA, USA, Paper Th4A.3, 2024.
[11] F. N. Sampaio et al., "Hollow-Core Fiber Transmission: Impact of $CO_2$ Absorption and its Mitigation by Waveform Design," *2025 European Conference on Optical Communications (ECOC)*, Copenhagen, Denmark, 2025, pp. 1-4.